\documentclass[conference]{IEEEtran}
\IEEEoverridecommandlockouts

\usepackage{url}
\usepackage{amsmath,amssymb,amsfonts}
\usepackage{algorithmic}
\usepackage{graphicx}
\usepackage{textcomp}
\usepackage{xcolor}
\usepackage{booktabs}
\usepackage{multirow}
\usepackage{siunitx}
\usepackage{tabularx}
\usepackage{ragged2e}
\usepackage{array}
\usepackage{cite}
\def\BibTeX{{\rm B\kern-.05em{\sc i\kern-.025em b}\kern-.08em
    T\kern-.1667em\lower.7ex\hbox{E}\kern-.125emX}}

\begin{document}

\title{The Sparsity Tax: Weight Sparsity Trade-offs in Event-Driven SIMD and SIMT Neuromorphic Cores}

\author{Mattias Westerink, Sameed Sohail, Berend-Jan van der Zwaag,\\ Sabih Gerez and Amirreza Yousefzadeh\\
University of Twente, Enschede, NL  
}

\maketitle

\begin{abstract}
Event-driven neuromorphic inference exploits activation sparsity by updating neuron state only on spikes. However, weight sparsity introduces irregular gather-style updates that undermine lockstep Single Instruction Multiple Data (SIMD) execution. We call the resulting overheads in control, metadata, and memory activity the sparsity tax. This paper quantifies that tax by comparing three closely related accelerators integrated into one neuromorphic core: (i) baseline lockstep SIMD, (ii) bitmap-gated Sparse-SIMD that selectively disables lanes without compressing weights, and (iii) a Single Instruction Multiple Threads (SIMT) style design with per-PE address generation and run-length coded sparse weights. Using an RTL-to-gates flow in GF22FDX+ and activity-driven energy estimation, we evaluate event-driven neural network inference across varying post-training pruning levels. Results show that total core area changes are insignificant because SRAM dominates area, while performance and energy strongly depend on how sparsity is handled: SIMD and Sparse-SIMD exhibit near-constant throughput, Sparse-SIMD achieves limited energy savings due to bitmap and dense-storage overheads, and SIMT provides the strongest energy scaling and substantial speedups at high sparsity, albeit with sublinear gains due to metadata reads, load imbalance, and sparsity-independent phases. The hardware code for the proposed architectures and experiments is publicly accessible for research purposes\footnote{\url{https://github.com/MW-SAND/neuromorphic-flos-co-processor/tree/main}}.

\end{abstract}

\section{Introduction}\label{sec:intro}

\begin{figure}
    \centering
    \includegraphics[width=1\linewidth]{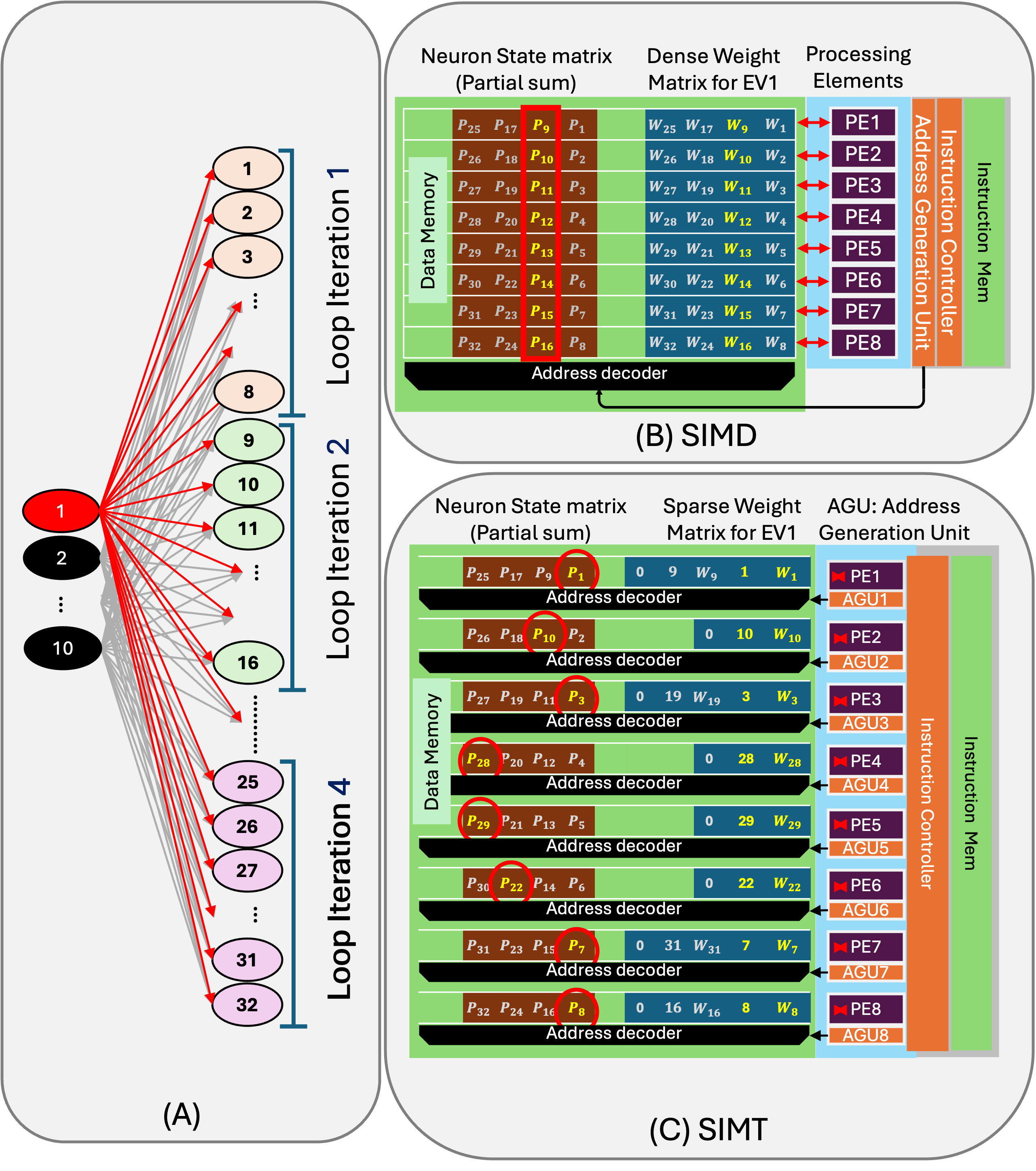}
    \caption{A) A layer of a neural network. When one input neuron activates and fires an event (EV1 in this figure), all output neurons must be updated, each with their corresponding weights (\( P_{i} \leftarrow P_{i} + W_{i} \times A_{i} \)), where \( P \) represents the partial sum and \( i \) is the index of the output neuron. B) An SIMD (Single Instruction, Multiple Data) processor features a shared control unit. The instruction controller fetches and decodes the instructions. The Address Generation Unit (AGU) generates data memory addresses for load/store instructions. C) A SIMT (Single Instruction, Multiple Threads) processor uses a sparse weight matrix and features individual AGUs for each processing thread. However, the instruction controller is shared, as all processing elements execute the same instructions.}
    \label{fig:sparse_weights_explain}
\end{figure}

Event-driven neural processing leverages activation sparsity by performing computations only when neurons fire, rather than updating all neurons at every time step \cite{xu2024optimizing, stuijt2021mubrain, moreira2020neuronflow, davies2018loihi, yousefzadeh2018hybrid, yousefzadeh2018performance, orchard2021efficient}. When an input event occurs, it typically triggers updates in multiple postsynaptic neurons, each requiring a multiply-accumulate (MAC) operation to adjust a running state or partial sum (e.g., \(P_i \leftarrow P_i + W_i \cdot A\)). As shown in Fig. \ref{fig:sparse_weights_explain}(A), a single active input can drive updates across several output neurons. These updates can be organized as a sequence of compact ``event kernels" that operate over contiguous neuron-state entries. Importantly, although the activity is temporally sparse, the per-event computation is often very regular, making it naturally amenable to data-parallel vectorization inside each neuromorphic core. These updates align neatly with a Single Instruction, Multiple Data (SIMD) organization (see Fig. \ref{fig:sparse_weights_explain}(B)) where a shared control unit broadcasts a single instruction stream, while multiple Processing Elements (PEs) execute the same operation on different neuron-state lanes.

However, this regularity breaks down when weight sparsity is introduced. Sparsifying weights has become increasingly common to reduce memory usage and avoid unnecessary computations \cite{gale2019state, han2016eie, zhou2018cambricon, parashar2017scnn}. When an event's outgoing weight vector is sparse, the update process becomes an irregular gather-accumulate operation: only a subset of output neurons needs to be updated, and these neurons may not be contiguous. In traditional SIMD architectures, leveraging this sparsity presents a challenge. The system is forced to either (i) maintain a dense representation, wasting energy by performing multiplications and accumulations (MACs) on zeros, or (ii) compress the weights, which incurs additional costs related to irregular control and addressing to skip the zeros. The main issue arises from the fact that decisions about skipping zeros and selecting target indices depend on individual processing elements. Each PE would require its own predicate or control mechanism to determine whether to participate in the computation or skip it. This situation undermines the primary advantage of SIMD, which is shared control that allows for uniform execution. It introduces what we call the ``sparsity tax": the extra energy, area, and latency costs associated with trying to regain efficiency in the presence of sparsity. 

Although several neuromorphic processors utilize SIMD to minimize controlling overhead \cite{tang2023seneca, gonzalez2024spinnaker2, narayanan2020spinalflow, lien2021vsa, agrawal2018spare, xu2024spiking}, to our knowledge, none of those effectively exploit unstructured weight sparsity. One effective method to address irregular sparsity is by adopting a SIMT (Single Instruction, Multiple Threads) execution model. In this model, each processing element functions like a lightweight thread with its own address generation and steering capabilities for each lane. Fig. \ref{fig:sparse_weights_explain}(C) illustrates this approach: the weights for events are stored in a sparse format that includes explicit indices and values. Each PE is equipped with a dedicated AGU (labeled AGU1-AGU8) to retrieve the correct neuron-state entry corresponding to the compressed weight. It's important to note that some control logic (instruction Controller in Fig. \ref{fig:sparse_weights_explain}) will still be shared, since all the PEs execute the same instructions for updating the neuron state with the incoming event. This organization enhances scalability in the presence of irregular sparsity, as the lanes can track distinct indices without requiring the entire vector to remain structured and dense. However, employing SIMT is not without its costs. Supporting per-thread address generation, metadata movement (e.g., indices), and the potential for execution imbalance can introduce overhead. These overheads may counteract or even outweigh the theoretical benefits of skipping zeros—especially in cases of moderate sparsity, when index widths are large relative to weight precision, or when memory energy consumption is more significant than the arithmetic operations involved. However, advanced neural network architectures such as Transformers can be extensively sparsified, with pruning of over 90\% of weights without significant accuracy loss \cite{zafrir2021prune}. Thus, leveraging weight sparsity in a neuromorphic processor has become essential. 

In this work, we quantify the trade-offs behind this sparsity tax by systematically comparing SIMD and SIMT neuromorphic core units under sparse event-driven workloads. Using the execution models in Fig. \ref{fig:sparse_weights_explain}(B) and Fig. \ref{fig:sparse_weights_explain}(C) as concrete architectural baselines, we provide fair area, power, and performance comparisons across sparsity regimes, isolating when sparsity is a net win and when its control and metadata costs dominate. Collectively, these results provide practical guidance for the design of next-generation event-driven processors.

\section{Proposed Architectures}

We propose three closely related accelerator architectures for event-driven neural-network updates that can be integrated into a neuromorphic core: an SIMD design, a Sparse-SIMD variant that introduces lightweight sparsity gating, and a SIMT variant that enables stronger sparsity exploitation by giving PEs greater autonomy in address generation. In this section, we will offer more detailed information about these accelerators.

\subsection{SIMD architecture}
 
The SIMD accelerator consists of a configurable array of processing elements that operate in an SIMD (Single Instruction, Multiple Data) fashion, with no dataflow between the PEs, as illustrated in Fig. \ref{fig:sparse_weights_explain}(B). Control is managed by an instruction controller, which selects the next instruction to execute (providing efficient branch and loop support) and translates each instruction into control signals for the PEs. 

The Address Generation Unit is responsible for preparing the proper addresses for data memory. Memory access occurs in lockstep. The memory is organized with a single wide port, meaning each read or write operation uses a single address. As a result, only one address calculator and one address decoder are utilized in the memory system. All PEs will access the same line of memory to perform SRAM operations. In theory, this allows data memory to be implemented using a single large SRAM block. However, in practice, multiple SRAM blocks are tiled together to create such memory, as a single large SRAM block is slow and not energy-efficient. 

\subsubsection*{Instruction Set Architecture (ISA) overview} 
The ISA is designed to cover the dominant operations in quantized neural network inference. It includes arithmetic instructions (Add, Mul, Max, Shift), where Max supports ReLU/pooling, and Shift supports common rescaling steps in integer-only quantized inference. Memory transfer uses Read/Write instructions that reference base addresses plus a compact immediate offset. A specialized Branch instruction executes instruction ranges N times using indexed loop counters, enabling compact loop nests for multi-dimensional layers.

\subsubsection*{Instruction Controller and Address Generation Unit} The Instruction Controller is programmed with tasks that define start and end instruction ranges, implementing nested loops sufficient for common neural network inference. Loop acceleration is implemented similarly to the loop controller in \cite{tang2023seneca}. For each memory access instruction, the Address Generation Unit determines the next memory address by adding the base address to a product of the loop indexes.

\subsubsection*{Processing Element}

\begin{figure} [!h]
    \centering
    \includegraphics[width=1\linewidth]{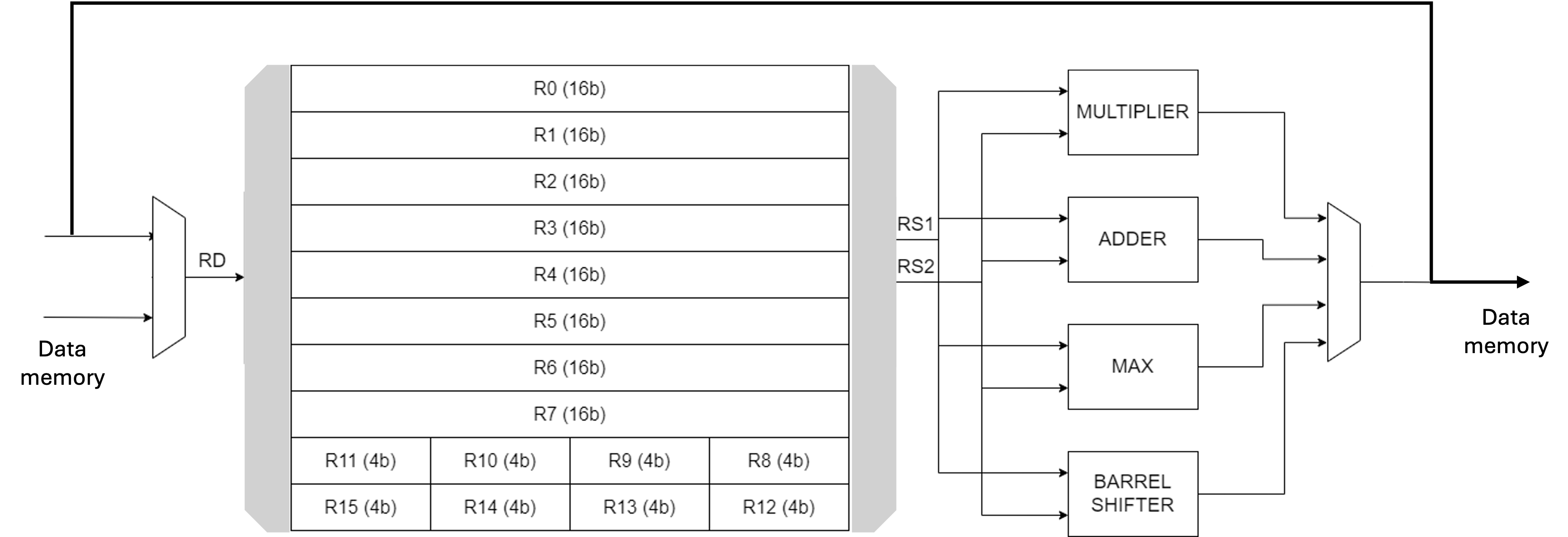}
    \caption{Architecture of the Processing Elements}
    \label{fig:pe}
\end{figure}

Processing elements operate in INT16 and INT4 formats, with a register file structured into 8$\times$16-bit and 8$\times$4-bit registers as illustrated in Fig. \ref{fig:pe}. Each PE is connected to the SRAM data memory through a dedicated 16-bit data read/write interface, sharing an address line with other PEs. Due to the 16-bit width of the SRAM data port, it can read four 4-bit weights simultaneously, minimizing weight-fetch overhead and eliminating the need for additional unpack instructions. Each PE can perform various operations, including addition, multiplication, finding the maximum value, and shifting. The modular arithmetic unit structure enhances extensibility. A zero-gating mechanism has been implemented to skip arithmetic operations when one of the operands is zero. This mechanism enables exploiting sparsity within the Arithmetic Logic Unit (ALU). However, it cannot eliminate the energy and time consumption associated with memory and register file access, which are dominant.

\subsection{Sparse-SIMD architecture (bitmap-gated SIMD)}

The baseline SIMD design struggles to effectively exploit unstructured weight sparsity because all Processing Elements share the same control signals and memory address stream. Consequently, a ``skip if weight == 0" approach cannot be applied to memory access. However, with minimal additional control, we have developed a Sparse-SIMD architecture that can dynamically disable individual PEs and their corresponding memory accesses for specific instructions issued by the instruction controller. This allows an instruction from the instruction controller to be executed in a few PEs while the others remain fully disabled. In this architecture, each PE can have its own dedicated portion of data memory (an SRAM row), and its port can be enabled or disabled independently. 

The key difference between Sparse-SIMD and regular SIMD is that instructions can be selectively disabled for individual PEs. For the PEs that do execute the instructions, they operate on the same long word of memory (one address, similar to the illustration in Fig. \ref{fig:sparse_weights_explain}(B)). This capability enables energy savings by eliminating redundant operations. However, this does not enhance performance, as skipping instructions simply turns off the PEs.

\subsubsection*{Bitmap-driven conditional execution} 
To disable memory access when weights are zero, we store sparsity metadata as a bitmap in data memory. Among encodings, a bitmap is attractive because its bits can be used directly as clock-enable signals for SRAM rows, registers, and arithmetic units, with relatively small control overhead. To avoid widening the instruction format, this design dedicates R7 as a “clock-enable register”. Operations involving R7 remain enabled, allowing the bitmap to be updated/shifted even under conditional execution. Therefore, R7 of each PE indicates the status (skip/execute) of the next 16 instructions and must be periodically updated with new data once every 16 instructions.

\subsubsection*{Microarchitectural changes} 
The instruction controller monitors whether conditional disablement is active and produces a clock-enable mode signal for the processing elements. When conditional disablement is active, each PE sends the least significant bit (LSB) of register R7 as an enable signal to its corresponding SRAM row and to the internal PE gating. When conditional disablement is inactive, the PE enable defaults to '1'. Additionally, the instruction controller temporarily overrides conditional disablement when an instruction targets R7, ensuring that the bitmap can be updated reliably. 

\subsubsection*{Implication}
\begin{itemize}
    \item \textbf{Memory footprint implication:} In this scheme, sparse weights are not stored in a compressed format. Consequently, zeros need to be stored in memory, just like in standard SIMD mode. Additionally, the bitmap used to disable certain instructions also consumes memory. Consequently, the overall memory consumption is always higher than in normal SIMD mode. However, the number of memory accesses is reduced for sparse weights.
    \item \textbf{Performance implication:} ``Bitmap-gated Sparse-SIMD" balances potential energy savings from skipping reads, writes, and computations on zero weights against the added instruction overhead of bitmap loading, shifting, and control. This overhead decreases throughput when sparsity is modest or when bitmaps require frequent refreshing.
\end{itemize}

\subsection{SIMT-style co-processor (per-PE address generation for sparse weights)}

To move beyond the basic ``enable/disable in lockstep" approach and fully leverage sparse formats that bypass runs of zeros, the third design introduces a SIMT-style extension. In this design, processing elements continue to execute a shared instruction stream, but each PE can utilize distinct base addresses and has its own dedicated address calculator, as illustrated in Fig. \ref{fig:sparse_weights_explain}(C). In SIMT execution, all PEs perform the same instructions (neuron updates) but operate on different addresses. However, the PEs are no longer aligned on the same loop iteration in Fig. \ref{fig:sparse_weights_explain}(A). This allows PEs to jump directly from one non-zero weight to another without pausing their operations, thereby enhancing throughput and energy efficiency, especially in high-sparsity scenarios. Finally, sparse weights can be stored in a compressed format, resulting in significant memory savings when sparsity is high.

\subsubsection*{Sparse encoding and ISA support} 
This design uses 4-bit run-length coding (RLC) to encode sparse matrices. In this method, we encode the number of zero weights (4b RLC) that follow each non-zero weight. Therefore, each non-zero 4b weight is now followed by a 4b RLC code. When sparsity exceeds 50\%, this encoding reduces the memory footprint. 

We chose 4-bit run-length coding (RLC) as a simple hardware-oriented sparse format that matches the INT4 weight precision used in this work and exposes the metadata/control trade-offs of interest without requiring complex pointer structures. The goal is not to claim that RLC is universally optimal, but to provide a concrete, sparse encoding for a fair architectural comparison. Other formats, such as CSR/CSC, coordinate encoding, or block sparsity, would change the metadata-to-weight ratio and address-generation behavior but would not change the central question studied here: whether sparse execution is better handled through lockstep gating or per-PE sparse traversal.

\subsubsection*{Address Generation Unit per PE} 
The main architectural addition is an AGU in each PE. The Address Generation Unit in each processing element uses the RLC value to compute the next partial sum address. Additionally, this unit can disable its corresponding PE and the SRAM blocks when there is no work to process.

\subsubsection*{Control-flow behavior under divergence} 
When per-PE enablement is active, some processing elements may finish their tasks earlier than others, while the remaining PEs continue to update their partial sums. Once all PEs have completed their tasks, the instruction controller will stop the current task and proceed to the next. This process ensures forward progress and re-convergence of operations.

\subsubsection*{Implication}
The SIMT-style extension shifts the focus of exploiting sparsity from lockstep coordination (as seen in Sparse-SIMD) to a more flexible approach. This method allows each Processing Element to navigate the sparse representation independently by updating its own addresses and enabling its own states. This approach offers two main advantages for sparse data: 
\begin{itemize}
    \item It only stores non-zero values, along with compact skip metadata, reducing the memory footprint.
    \item It allows for skipping both computations and partial-sum memory access/operations. This can significantly enhance energy efficiency and throughput.
\end{itemize}

However, this method comes with increased microarchitectural overhead. Notably, it requires a per-PE AGU, additional base-address states, and mechanisms to manage divergence and reconvergence. These factors can lead to increased area usage, higher static power consumption, and potential control stalls (such as re-enable or branch overheads) when the skip fields are short or irregular.

As a result, SIMT generally outperforms lockstep designs in high, irregular sparsity, where the benefits of skip metadata can be fully realized. Conversely, its advantages diminish in dense or lightly sparse layers, where the overhead associated with per-PE addressing and control can approach the costs of simply executing the standard SIMD stream.

Table~\ref{tab:microarch_summary} summarizes the key microarchitectural differences between the three accelerator variants in terms of control, addressing, sparse-data representation, and execution behavior.

\begin{table}[t]
\centering
\caption{Microarchitectural comparison of the three accelerator variants.}
\label{tab:microarch_summary}
\renewcommand{\arraystretch}{1.08}
\setlength{\tabcolsep}{3pt}
\footnotesize
\begin{tabularx}{\columnwidth}{@{}p{2.45cm} >{\centering\arraybackslash}p{1.25cm} >{\centering\arraybackslash}p{1.75cm} >{\centering\arraybackslash}X@{}}
\toprule
\textbf{Feature} & \textbf{SIMD} & \textbf{Sparse-SIMD} & \textbf{SIMT} \\
\midrule
Instruction stream   & Shared     & Shared         & Shared \\
Address generation   & Shared  & Shared AGU     & Per-PE AGU \\
Weight storage       & Dense      & Bitmap         & Sparse RLC \\
Per-PE enable        & No         & Bitmap-driven  & Autonomous \\
Skip zero updates    & No         & Partial        & Yes \\
Sparse work skipping & No         & No             & Yes \\
Main overhead        & Dense zeros & Bitmap overhead & Metadata + divergence \\
\bottomrule
\end{tabularx}
\end{table}

\section{Experimental results}


\subsection{Experimental Setup}

\subsubsection*{System architecture and RISC-V}


To evaluate the proposed SIMD, Sparse-SIMD, and SIMT extensions in a realistic control setting (without requiring a custom software stack), we integrated these accelerators into a platform built around a NEORV32 \cite{neorv} RISC-V processor.
The RISC-V core acts as a lightweight runtime controller that (i) parses input-event packets, (ii) programs accelerator base addresses and instructions, and (iii) triggers accelerator tasks for partial-sum updates and activation/quantization steps. This setup allows all experiments to be driven from compact C programs compiled for the embedded RISC-V.

We used the clock frequency of 500 MHz for our experiments and included (in addition to the RISC-V and accelerator) banked SRAM-based instruction and data memory that stores runtime variables, network weights, and intermediate partial sums. The accelerator is logically loosely coupled to NEORV32 through memory-mapped registers, while completion signaling is handled via the NEORV32 FIRQ\_CFS interrupt mechanism. In contrast, the accelerator is tightly coupled to the data memory: it directly drives the SRAM control signals and exchanges a wide data interface of P×16 bits (P is the number of processing elements). When the NEORV32 and accelerator contend for data-memory access, the accelerator is prioritized, and the NEORV32 is stalled, ensuring deterministic accelerator progress. 

\subsubsection*{Instruction and Data memories}

The instruction and data memories are implemented using single-port, ultra-high-density, low-voltage-enabled SRAMs generated with Synopsys Embed-It Integrator. The instruction memory consists of 8 SRAM macros of 32 bits × 1024 words (8×4KB). The data memory consists of 64 SRAM macros of 16 bits × 2048 words (64×4KB). These configurations were chosen as a practical trade-off between memory density and energy consumption.

\subsubsection*{Technology node, Simulation and measurement tools}

We used an RTL-to-gates evaluation flow to report cycle counts and activity-driven power. Functional simulation and cycle measurements were performed in Cadence Xcelium (23.03.007). Cycle counts were collected using the NEORV32 hardware performance monitor (HPM) counters while running the compiled benchmark code on the simulated platform. All experimental kernels were written in C and compiled using the riscv-none-elf toolchain as described in the NEORV32 user documentation, enabling consistent software control across SIMD/Sparse-SIMD/SIMT variants.

Synthesis and area extraction were performed using Cadence Genus (23.33) and GF22FDX+ LVT20 ultra-low-power libraries. Energy consumption was estimated in Cadence Joules (23.33) using activity-driven analysis based on switching activity generated by simulation, reported to be accurate within 15\% of signoff power. This approach captures the effect of sparsity-dependent memory and datapath enablement on dynamic power, which is essential for comparing lockstep (SIMD/Sparse-SIMD) and per-PE control (SIMT) designs across varying sparsity levels. 

\subsection{Benchmarking application (event-based VGG-16)}

\begin{table}[!h]
    \centering
    \caption{Customized VGG-16 layer list and baseline SIMD latency/energy per layer. Each row corresponds to one layer of the network. Convolution kernels are $3\times3$.}
    \begin{tabular}{|c|c|c|c|c|}
        \hline 
                      &           &             & SIMD    & SIMD  \\
        Name          & Output    & Activation  & Latency & Energy\\
                      & Shape     & Sparsity    & (ms)    & (mJ)  \\
        \hline \hline
        Input         & 32,32,3   &     0       &   -     &   -   \\ \hline 
        block1\_conv1 & 32,32,64  &    1.6\%    &  15.6   &  1.96 \\ \hline 
        block1\_conv2 & 32,32,64  &   47.7\%    &  174    &  21.9 \\ \hline 
        block1\_pool  & 16,16,64  &   -         &   -     &   -   \\ \hline 
        block2\_conv1 & 16,16,128 &   24.8\%    &  241    &  15.8 \\ \hline 
        block2\_conv2 & 16,16,128 &   38.7\%    &  393    &  25.7 \\ \hline 
        block2\_pool  & 8,8,128   &   -         &   -     &   -   \\ \hline
        block3\_conv1 & 8,8,256   &   39.7\%    &  358    &  12.6 \\ \hline 
        block3\_conv2 & 8,8,256   &   58.1\%    &  436    &  17.6 \\ \hline 
        block3\_conv3 & 8,8,256   &   58.7\%    &  430    &  17.3 \\ \hline 
        block3\_pool  & 4,4,256   &   -         &   -     &   -   \\ \hline
        block4\_conv1 & 4,4,512   &   64.3\%    &  247    &  7.48 \\ \hline 
        block4\_conv2 & 4,4,512   &   74.7\%    &  263    &  10.6 \\ \hline 
        block4\_conv3 & 4,4,512   &   85.4\%    &  152    &  6.13 \\ \hline 
        block4\_pool  & 2,2,512   &    -        &   -     &   -   \\ \hline
        block5\_conv1 & 2,2,512   &   79.4\%    &  71.6   &  2.16  \\ \hline 
        block5\_conv2 & 2,2,512   &   87.4\%    &  43.8   &  1.32 \\ \hline 
        block5\_conv3 & 2,2,512   &   88.5\%    &  39.9   &  1.21 \\ \hline
        block5\_pool  & 1,1,512   &    -        &   -     &   -   \\ \hline 
        dense         & 512       &    98\%     &  9e-3   &  6e-5 \\ \hline 
        dense\_1      & 256       &    98\%     &  5e-3   &  3e-5 \\ \hline 
        Output        & 10        &    -        &  -      &  -    \\ \hline 
    \end{tabular}
    \label{tab:VGG-16_architecture}
\end{table}

The design is evaluated using the VGG-16 neural network architecture \cite{VGG} customized for the CIFAR-10 dataset \cite{CIFAR}. This enables the comparison with other accelerators that provide layer-wise results for VGG-16. The deployed network has undergone quantization-aware training, with weights and activations quantized to 4 bits each.

Table \ref{tab:VGG-16_architecture} lists all evaluated layers of the customized VGG-16 model and reports the baseline SIMD latency and energy for each layer. The model is constructed with Tensorflow 2.14.0.

We compared memory footprint, execution time (throughput), and energy consumption across three variations: baseline SIMD, bitmap-gated Sparse-SIMD, and SIMT, using ``VGG16" layers \cite{VGG} driven by event packets. Each event carries the information about the source neuron address (the neuron from the previous layer that generated the spike) and its corresponding activation (assuming graded spikes). The reported measurements include address generation from the input events, updates to partial sums, and the quantization/activation step. However, they exclude address generation and serialization of output-event packets.

To induce weight sparsity in the neural network, we prune the smallest weights using a variable threshold (post-training). Therefore, we expect a decrease in accuracy as thresholds increase. However, this accuracy loss can be [partially] recovered when using advanced pruning techniques \cite{zhu2025comprehensive}. We omit advanced retraining-aware pruning techniques because our focus is architectural. The choice of thresholding method vs more advanced methods doesn’t affect the architectural trends.

\subsection{Area breakdown}

\begin{table*}[!t]
\centering
\caption{Area breakdown of a neuromorphic core comparing SIMD, Sparse-SIMD, and SIMT accelerators (8 processing elements). [kµm² = 10³ µm²]}
\label{tab:nc_area}
\setlength{\tabcolsep}{6pt}
\renewcommand{\arraystretch}{1.15}

\begin{tabular}{
  l
  S[table-format=3.2]
  S[table-format=3.2]
  S[table-format=3.2]
  l
}
\toprule
\textbf{Component} &
\textbf{SIMD} &
\textbf{Sparse-SIMD} &
\textbf{SIMT} &
\textbf{Overhead vs. SIMD} \\
& \textbf{(\si{k\micro\meter\squared})} &
  \textbf{(\si{k\micro\meter\squared})} &
  \textbf{(\si{k\micro\meter\squared})} & \\
\midrule

\textbf{Total}                   & 528   & 529   & 529   & Sparse-SIMD/SIMT: +0.2\% \\

\addlinespace[2pt]
NEORV32 core                     & 7.25  & 7.25  & 7.25  & 0\% \\
256\,KB data memory              & 452   & 452   & 452   & 0\% \\
32\,KB instruction memory        & 56.6  & 56.6  & 56.6  & 0\% \\

\addlinespace[2pt]
\textit{\textbf{Accelerator}} \\
Accelerator (total)              & 11.8  & 12.9  & 12.9  & Sparse-SIMD/SIMT: +9.3\% \\
\quad 8 processing elements      & 7.25  & 7.46  & 7.30  & Sparse-SIMD: +2.9\%, SIMT: +0.7\% \\
\quad Instruction controller     & 4.11  & 4.97  & 4.11  & \textbf{Sparse-SIMD: +20.9}\%, SIMT: 0\% \\
\quad Address Generation Unit(s)     & 0.30  & 0.30  & 1.21  & Sparse-SIMD: 0\%, \textbf{SIMT: $\times$4.0} \\
\bottomrule
\end{tabular}
\end{table*}

Table \ref{tab:nc_area} summarizes the post-synthesis area of a single neuromorphic core based on NEORV32, comparing a baseline core (“SIMD”) against Sparse-SIMD, and SIMT co-processors (all with 8 PEs).

The small overall area delta is explained by the fact that data memory dominates the core area (85–88\%), with instruction memory contributing another 10–11\%; these components remain essentially constant across all variants. Consequently, the net increase is primarily due to the accelerator's logic. Inside the accelerators, the area is dominated by their 8 Processing Elements. Relative to the SIMD accelerator:
\begin{itemize}
    \item Sparse-SIMD increases accelerator area by 9.24\%, largely due to Instruction-Controller growth. 
    \item SIMT increases accelerator area by 8.88\%, primarily from adding per-PE Address Generation Units. 
\end{itemize}
Therefore, SIMT’s total accelerator area (12,862 µm²) is almost the same as the Sparse-SIMD (12,905 µm²), because the added AGU area is approximately offset by a smaller Instruction Controller in SIMT.

\subsection{Memory footprint overhead}

\begin{figure} [!h]
    \centering
    \includegraphics[width=0.99\linewidth]{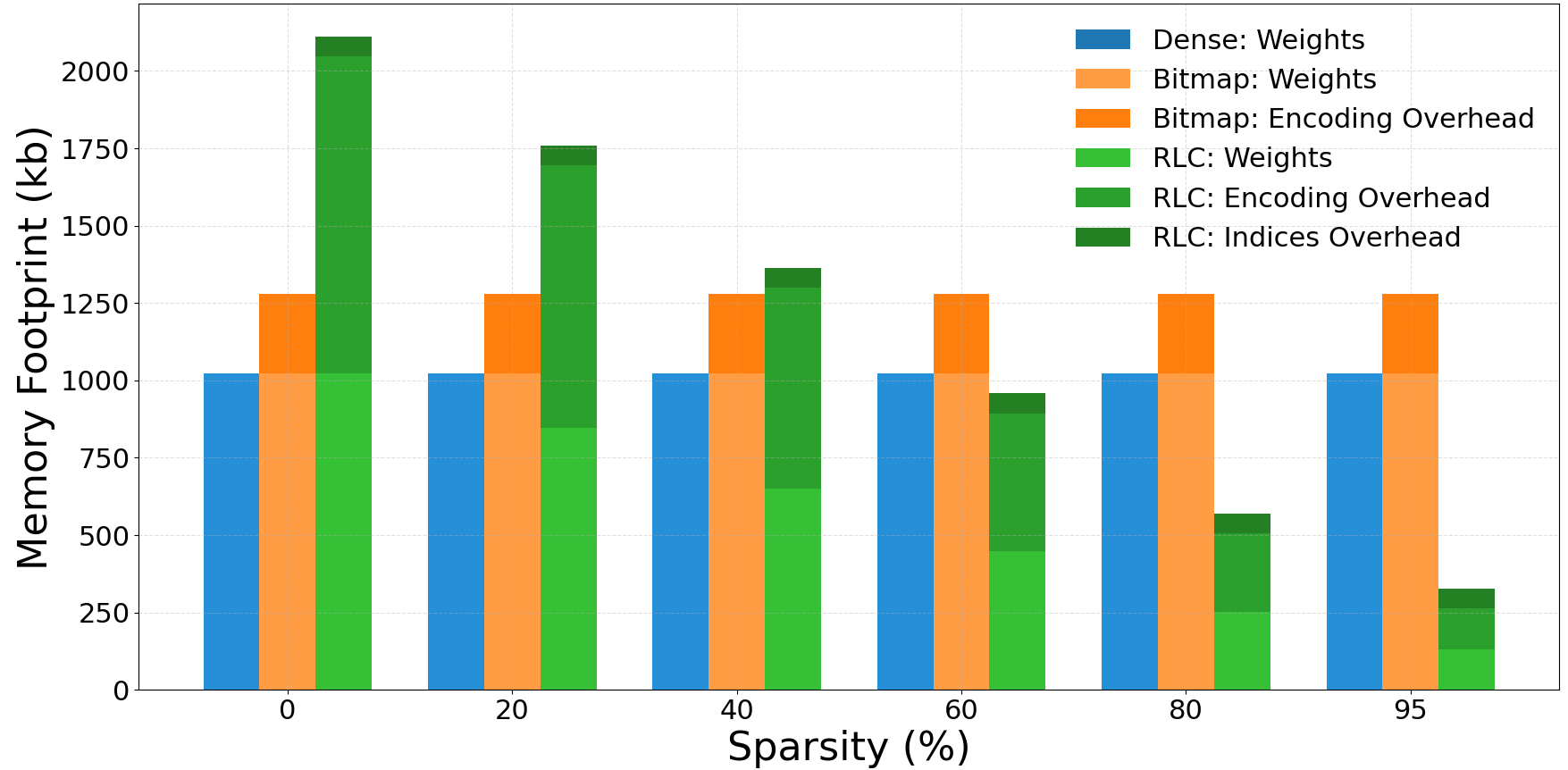}
    \caption{Memory Footprint of ``VGG-16 dense layer" Weights for SIMD (Dense storage, in Blue), Sparse-SIMD (Bitmap storage, in Orange) and SIMT (RLC Encodings storage, in Green) at Six Levels of Sparsity. Sparsity is achieved by reducing smaller weights to zero.}
    \label{fig:m10_footprint_fc}
\end{figure}

The architecture and its corresponding weight encoding affect the memory footprint. The memory footprint for the ``VGG-16 dense" layer is reported in Fig. \ref{fig:m10_footprint_fc}. The relative memory footprint of the encodings is roughly the same across all layers of VGG-16: 
\begin{itemize}
    \item SIMD architecture: In this case, the dense encoding is used with 1024kb for 262144 (512×512) Int4 weights.
    \item Sparse-SIMD architecture:  In this case, the bitmap encoding is used with 1024 $kb$ for the weights and 256 $kb$ for the bitmaps. Since the weights are not stored in compressed sparse mode, the sparsity does not affect the memory footprint. 
    \item SIMT architecture: In this case, Run Length Encoding (RLC) is used, and the weights are stored in compressed sparse format. The encoding overhead of the inserted RLC values equals the number of stored weights. As such, the overhead of the RLC encoding at 0\% sparsity is 106.25\%. The last 6.25\% comes from the stored indices (64kb). 
    
\end{itemize}
 
At 95\% sparsity, RLC encoding reduces the memory footprint by 67.83\%. Ideally, this reduction should be around 83.75\%, including index overhead. This discrepancy occurs because a single 4-bit RLC value can represent the skipping of up to 15 intermediate zero weights. However, for larger jumps, one or a few intermediate zero-value weights need to be stored to limit the address jump. The lower reduction in memory observed is mainly due to the inclusion of these zero-valued weights.

\subsection{Performance comparison}

\begin{figure}
    \centering
    \begin{minipage}{\columnwidth}
    \includegraphics[width=0.99\textwidth]{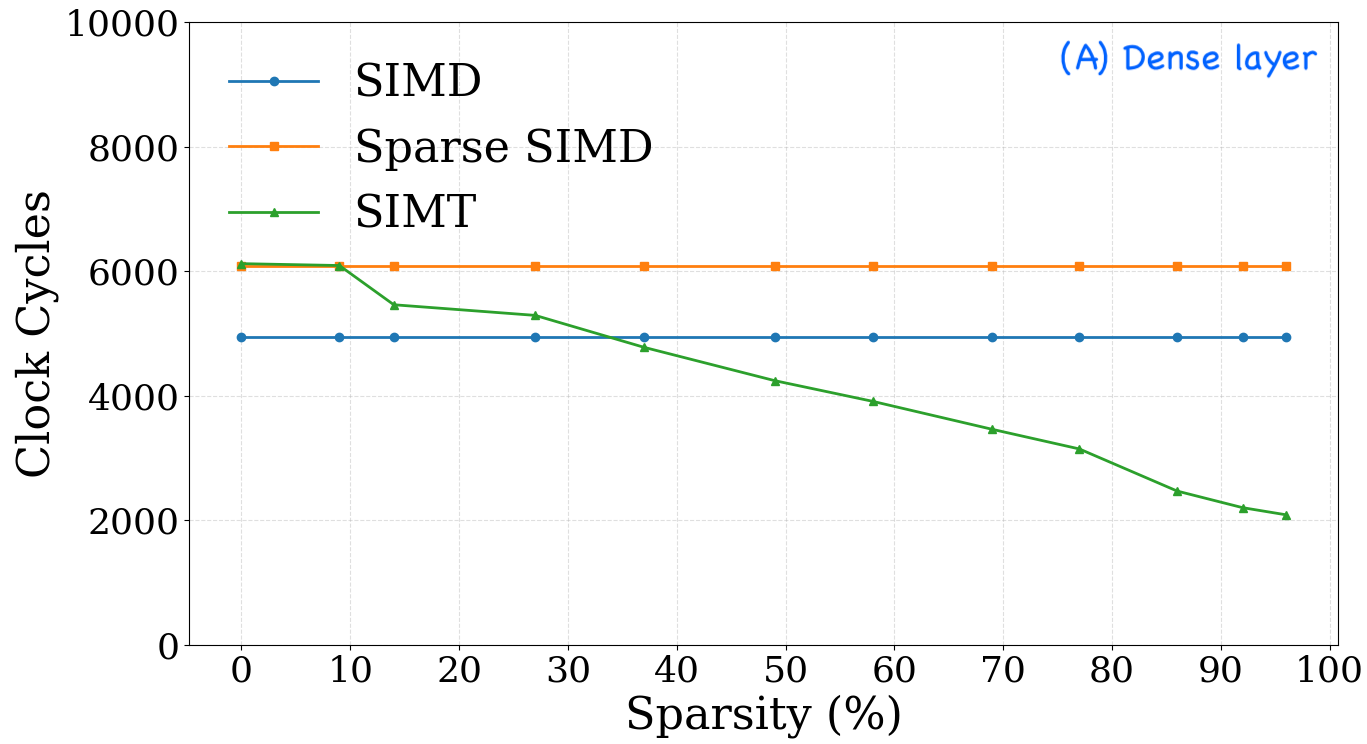}
    \end{minipage}\hfill
    \begin{minipage}{\columnwidth}
    \includegraphics[width=0.99\textwidth]{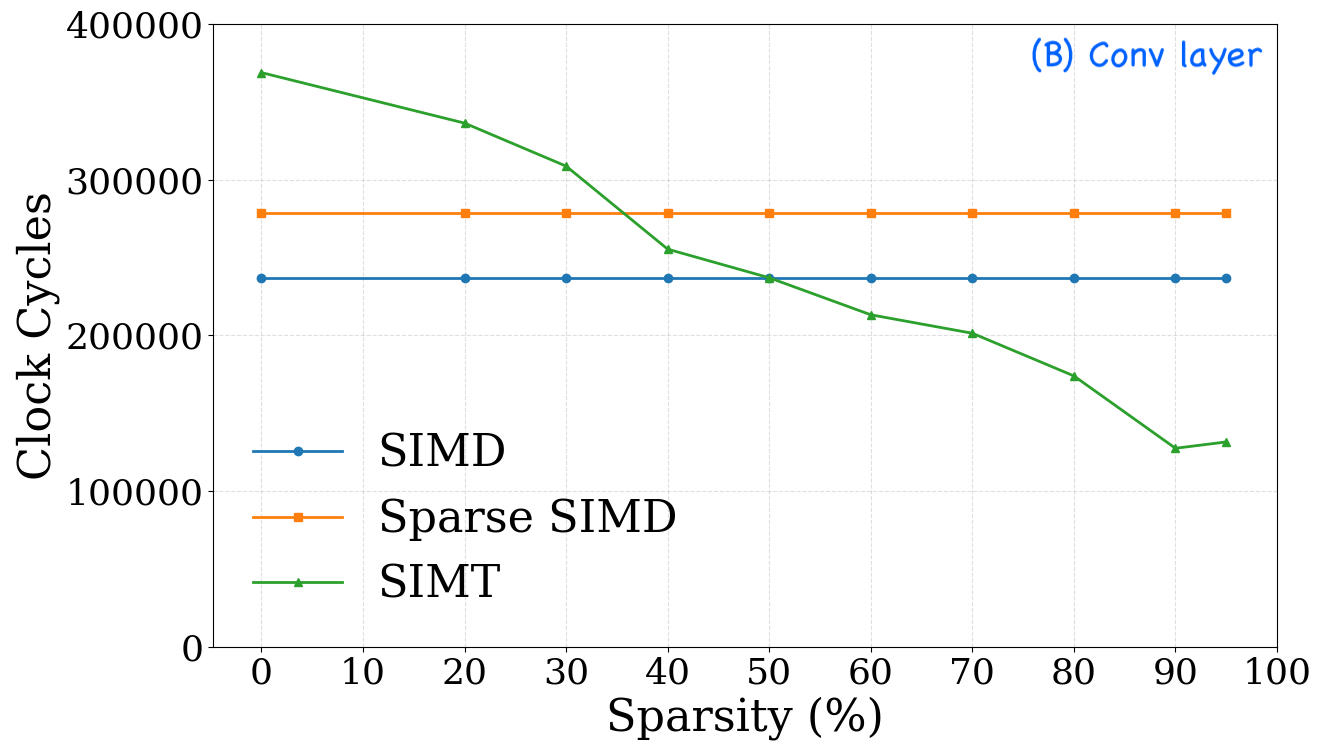}
    \end{minipage}\hfill
    \caption{Effect of Sparsity on Execution Time (in clock cycles) for SIMD, Sparse-SIMD and SIMT, evaluated with the (a) VGG-16 dense Layer, (b) VGG-16 block2 conv1 Layer. The clock cycles shown are the total cycles required to process the layer, using an input-event count derived from the average activation sparsity across CIFAR-10 images.}
    \label{fig:clock-cycles-sparsity}
\end{figure}

Fig. \ref{fig:clock-cycles-sparsity} illustrates the number of clock cycles required to process both a dense layer and a convolutional layer in VGG-16 across all three accelerator architectures. The throughput of the SIMD and Sparse-SIMD accelerators remains constant and is independent of the weight sparsity. This is expected since neither architecture bypasses computations involving zero-valued weights. 

The clock cycle count for the Sparse-SIMD accelerator is more than that of the SIMD accelerator. This increase is anticipated because its instruction set incurs the overhead of fetching bitmaps and performing a shift operation on them to activate or deactivate the execution of subsequent instructions.

As expected, the execution time of the SIMT architecture decreases gradually as sparsity increases. This can be explained by the skipping of computations and data transfers with zero-valued weights. 

For low sparsity levels, the SIMT unit incurs overhead, resulting in worse performance than SIMD. This is similar to the Sparse-SIMD architecture. However, in convolutional layers, SIMT overhead is significantly higher because our SIMT architecture fails to exploit the potential reduction in operations at the edges of the output feature map. To do so, it would need to know the starting point of each column's weight set. However, since the weights are stored in RLC encoding, calculating the start point is not straightforward. It's important to note that this limitation is a design choice and should not be generalized to all SIMT architectures.

At the highest level of weight sparsity, the SIMT accelerator achieves more than a 50\% reduction in execution time compared to the SIMD accelerator. However, since 96\% of the partial sum updates need to be skipped, the actual reduction in execution time is much smaller than the increase in sparsity might suggest. Several factors contribute to this diminished reduction in execution time:
\begin{itemize}
    \item The execution time of the SIMT accelerator increases due to the additional instruction required for each partial sum update.
    \item For every 4-bit weight, a 4-bit RLC value must be read, effectively doubling the memory accesses for each non-zero weight.
    \item Variations in sparsity lead to some Processing Elements being idle while others are active, resulting in an uneven workload distribution.
    \item The execution time of some tasks, such as the event generation task (applying activation functions to the partial sums), is not affected by sparsity and remains fixed. 
\end{itemize}
These factors combined limit the overall efficiency gains from increased sparsity in the SIMT accelerator. 

\subsection{Energy consumption comparison}

\begin{figure} [!h]
    \centering
    \includegraphics[width=0.99\linewidth]{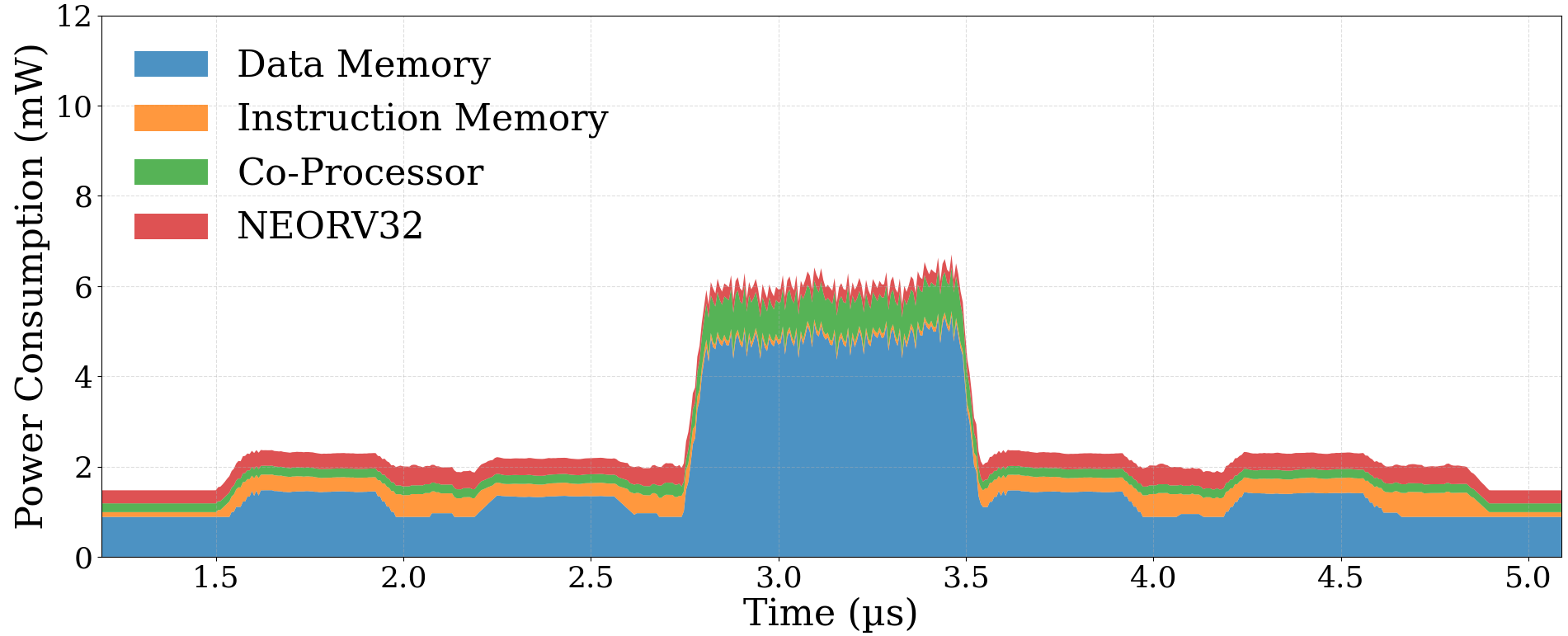}
    \caption{Distribution of power consumption over Time for processing a Single Input Event in the dense layer with SIMD accelerator (8 PEs)}
    \label{fig:m8_power_over_time}
\end{figure}

\begin{figure}
    \centering
    \begin{minipage}{\columnwidth}
    \includegraphics[width=0.99\textwidth]{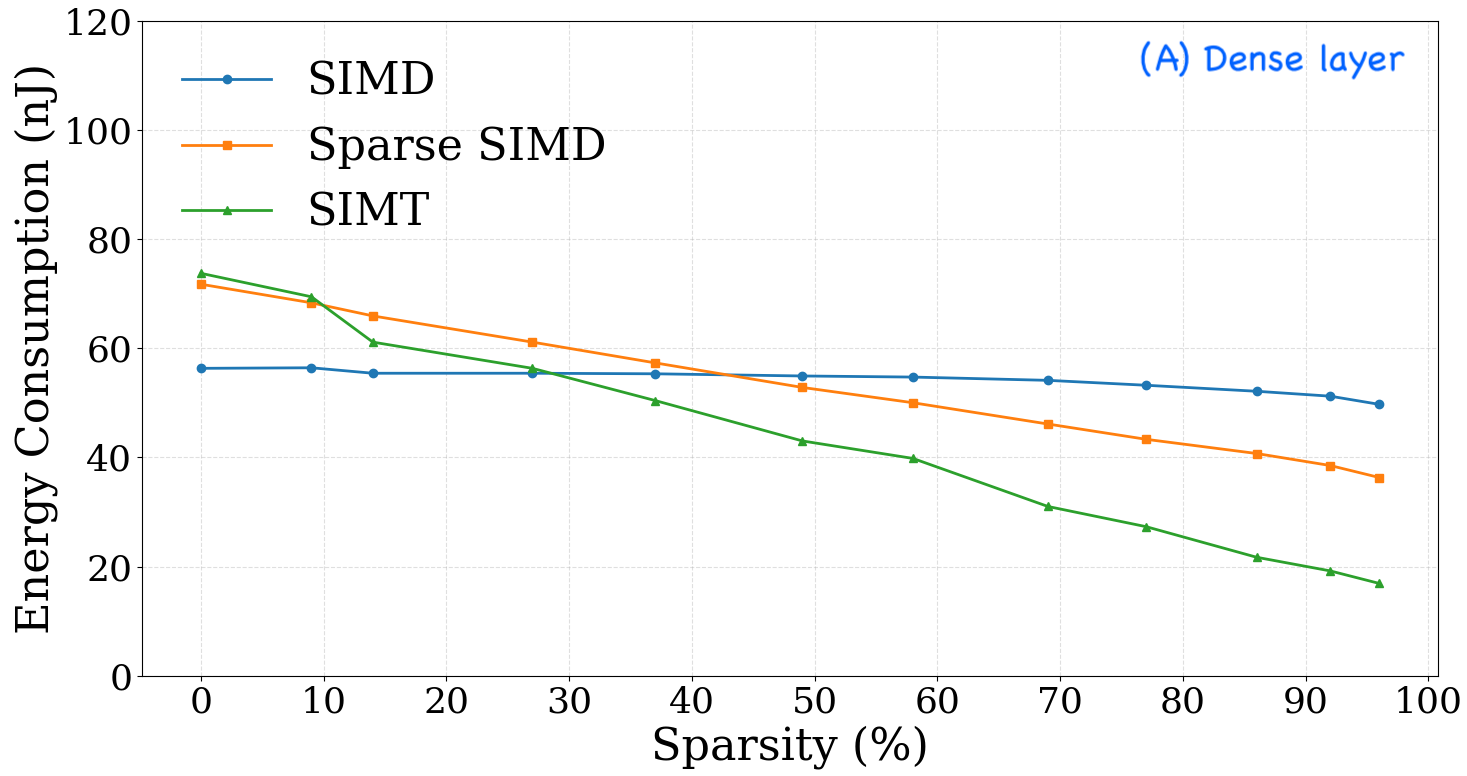}
    \end{minipage}\hfill
    \begin{minipage}{\columnwidth}
    \includegraphics[width=0.99\textwidth]{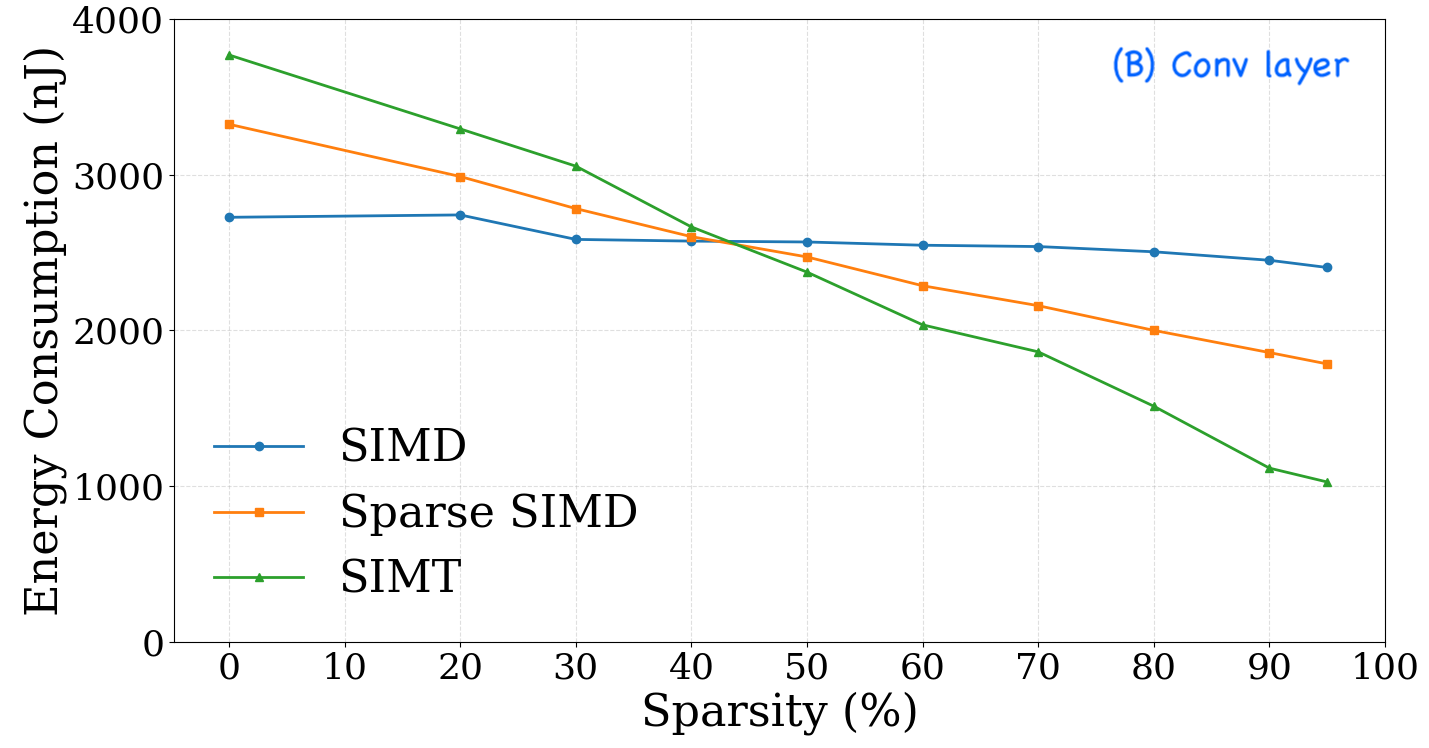}
    \end{minipage}\hfill
    \caption{Effect of Sparsity on energy consumption for SIMD, Sparse-SIMD and SIMT, evaluated with the (a) VGG-16 dense Layer, (b) VGG-16 block2 conv1 Layer. The energy consumption shown is the total amount of energy required to process the layer, using an input-event count derived from the average activation sparsity across CIFAR-10 images.}
    \label{fig:energy_sparsity}
\end{figure}

\begin{table*}[t]
\centering
\caption{Comparison of several neuromorphic cores concerning event data type, performance, area, and energy consumption. Measurements are conducted with a 4000-neuron fully connected layer featuring 16 input events at 86\% weight sparsity.}
\label{tab:neuromorphic-cores-comparison}
\setlength{\tabcolsep}{2pt}
\renewcommand{\arraystretch}{1.12}

\begin{tabular}{l l c c
                S[table-format=1.2]
                S[table-format=1.2]
                S[table-format=2.0]
                S[table-format=1.1]}
\toprule
\textbf{Accelerator} &
\multicolumn{1}{c}{\textbf{Architecture}} &
\multicolumn{1}{c}{\textbf{Technology}} &
\multicolumn{1}{c}{\textbf{Event}} &
\multicolumn{1}{c}{\textbf{Area per}} &
\multicolumn{1}{c}{\textbf{Memory per}} &
\multicolumn{1}{c}{\textbf{Energy per}} &
\multicolumn{1}{c}{\textbf{Time per}} \\
& \multicolumn{1}{c}{\textbf{or Mode}} &
& \multicolumn{1}{c}{\textbf{Datatype}} &
\multicolumn{1}{c}{\textbf{core (mm$^{2}$)}} &
\multicolumn{1}{c}{\textbf{core (Mb)}} &
\multicolumn{1}{c}{\textbf{neuron update (pJ)}} &
\multicolumn{1}{c}{\textbf{neuron update (ns)}} \\
\midrule
Loihi\cite{davies2018loihi}      & One PE per core & 14 nm & Binary   & 0.41 & 2.0  & 81 & 8.4 \\
\midrule
NeuronFlow\cite{moreira2020neuronflow} & One PE per core & 28 nm & Binary   & 0.10 & 0.12 & 20 & NA \\
\midrule
\multirow{2}{*}{SENECA\cite{tang2023seneca}} &
SIMD8-BF16 & 22 nm & BF16 & 0.47 & 2.3 & 14 & 3.6 \\
& SIMD8-INT8 & 22 nm & Binary & 0.47 & 2.3 & 7  & 1.7 \\
\midrule
\multirow{3}{*}{\textbf{Our Accelerators}} &
SIMD8         & 22 nm & INT4 & 0.53 & 2.3 & 10 & 1.7 \\
& Sparse-SIMD8 & 22 nm & INT4 & 0.53 & 2.3 & {7} & {2.1} \\
& SIMT8        & 22 nm & INT4 & 0.53 & 2.3 & {4} & {0.9} \\
\bottomrule
\end{tabular}

\vspace{1.5mm}
\end{table*}

Fig. \ref{fig:m8_power_over_time} illustrates the instantaneous power consumption of the SIMD configuration while processing a single input event. The baseline (idle) power consumption is approximately 1.5 mW, primarily due to leakage and standby activity in the instruction and data memories. When the NEORV32 performs instruction fetches and occasional data accesses, power consumption increases to about 2–2.5 mW. Once the accelerator is activated, power rises to around 5.5–6 mW, experiencing brief peaks of up to 6.5 mW. This data confirms that data memory activity is the main contributor to dynamic power consumption during event processing, while the processing element array and control logic contribute a smaller fraction of the total, consistent with the area breakdown, indicating that SRAM usage is predominant. The short power peak around 3µs corresponds to the active partial-sum update phase, during which the accelerator is enabled, and the data memory experiences the highest switching activity due to repeated weight/partial-sum accesses.

Fig. \ref{fig:energy_sparsity} compares the total energy per layer (including event parsing, partial-sum updates, and activation/quantization) for SIMD, Sparse-SIMD, and SIMT across increasing levels of weight sparsity, evaluated on both (a) a dense layer and (b) a convolutional layer. Several trends are observed:
\begin{itemize}
    \item SIMD: small energy reduction despite dense execution. While baseline SIMD does not skip over zeros at the memory-access level, its energy consumption decreases slightly as sparsity increases. This is because zeros reduce switching activity: fewer bit toggles travel through the SRAM read datapath, the register file, and the arithmetic units, especially with zero-gating implemented in the ALU. Consequently, this lowers dynamic power usage, even when executing the same instruction stream.
    \item Sparse-SIMD: energy decreases with sparsity, but with a limited slope. Bitmap-gated Sparse-SIMD can disable subsets of PEs (and their associated SRAM accesses) using a per-lane clock enable derived from the bitmap. As sparsity increases, more lanes are disabled more often, cutting dynamic energy in the PE datapath and partial-sum/weight memory accesses. However, the achievable energy reduction is tempered by the bitmap mechanism, which introduces overhead instructions (bitmap fetch/shift/refresh) that do not vanish with sparsity and therefore impose a non-negligible constant energy component.
    \item SIMT: strongest energy scaling with sparsity. SIMT demonstrates the most significant reduction in energy consumption as sparsity increases because it simultaneously addresses two major costs: the amount of work done and memory activity. By utilizing a compressed sparse representation (RLC + values), SIMT achieves the following: (i) it completely skips zero-weight elements, thereby avoiding both multiply-accumulate (MAC) operations and the corresponding memory traffic associated with these skipped updates, and (ii) it reduces switching by issuing fewer effective updates as sparsity increases. Furthermore, unlike Sparse-SIMD, which primarily saves energy by controlling power within a fixed-length stream, SIMT also decreases the overall work executed at the layer level. This compounded effect results in greater energy savings when sparsity is high.
\end{itemize}

To summarize, Sparse-SIMD saves energy per instruction by gating PEs and SRAM rows, but it still incurs overhead from a largely fixed control schedule that contains non-skippable segments. SIMT, on the other hand, conserves energy by avoiding the execution of the skipped updates altogether. 

As shown, the benefits of SIMT are not perfectly proportional to sparsity levels. This is due to the introduction of certain overheads, such as additional metadata reads (like RLC values), as well as divergence and imbalance among processing elements. Moreover, there are phases that are independent of sparsity, such as activation and event generation.

\subsection{Comparison with existing neuromorphic processors}

Although the primary goal of this work is an apples-to-apples comparison between three closely related co-processors integrated in the same NEORV32-based core, it is still useful to position the resulting operating points against representative digital neuromorphic systems reported in prior work. Table \ref{tab:neuromorphic-cores-comparison} summarizes a normalized comparison using common benchmarks. 

These numbers suggest that, for this workload, our SIMT design achieves the lowest per-update energy and latency among the compared points. At the same time, the comparison should be interpreted cautiously because the platforms differ in event representation (binary vs graded), precision formats, memory organization, and measurement methodology, all of which can materially affect the reported pJ/update and ns/update. 

Many neuromorphic processors typically feature only one Processing Element per core \cite{davies2018loihi, moreira2020neuronflow, orchard2021efficient, akopyan2015truenorth, furber2014spinnaker}. These processors often incur significant control overhead, resulting in poorer performance than SIMD or SIMT cores.

\section{Conclusion}

This paper quantifies the sparsity tax in event-driven neuromorphic processing by comparing three architectures: the baseline lockstep SIMD, the bitmap-gated Sparse-SIMD, and the SIMT with per-PE address generation, all within the same NEORV32-based core, SRAM-dominated memory system, and synthesized on GF22nm-FDX+ technology. With eight PEs, the differences in total core area across these architectures are insignificant.

The results distinguish between gating sparsity and skipping sparsity, revealing how each affects both performance and energy consumption. SIMD and Sparse-SIMD maintain nearly constant throughput regardless of sparsity because they primarily execute the same workload (though Sparse-SIMD incurs a fixed overhead from bitmap usage). In contrast, SIMT performance improves as sparsity increases, as it effectively traverses compressed representations and skips over zeros. However, the performance gains are sublinear due to factors such as metadata reads, PE divergence or imbalance, and phases of computation independent of sparsity.

Energy consumption follows a similar pattern: SIMD experiences only slight reductions in energy usage due to lower switching activity. Sparse-SIMD saves more energy by disabling subsets of PEs and their associated memory accesses, though it is limited by the costs associated with densely packed zeros and bitmap overhead. SIMT offers the largest energy savings by avoiding both computation and memory traffic for skipped zeros, though this is offset by the costs of metadata movement and potential imbalances across PEs.

The quantitative break-even point presented in this work should be understood as specific to our evaluated context, which includes INT4 event-driven inference, VGG-16-style workloads, and 4-bit RLC sparse encoding. In practice, the exact transition between SIMD-style execution and SIMT-style sparse traversal will vary based on factors such as weight precision, metadata format, sparsity structure, and workload type. For instance, higher precision reduces the relative cost of metadata, while different sparse encodings or more structured sparsity may affect both address-generation overhead and lane divergence. Nonetheless, the qualitative conclusion is expected to remain valid across a broader scope: lockstep gating primarily conserves energy within a largely fixed execution schedule, whereas architectures that effectively skip zero-weight updates can also reduce total workload, making them increasingly appealing as sparsity becomes more irregular.

Overall, SIMD is appealing for dense or lightly sparse workloads. Sparse-SIMD offers a low-disruption option for energy optimization when the storage format cannot be altered, and SIMT proves most effective for high, irregular sparsity, especially if future research can reduce metadata overhead and improve load balancing and handling of sparse layouts, such as convolution boundaries. When the weight sparsity exceeds 50\%, which is easily achievable in advanced neural networks, the sparsity tax becomes almost negligible, and its advantages become evident.

\section*{Acknowledgment}
We kindly acknowledge EUROPRACTICE for its support with the design tool.

\bibliographystyle{IEEEtran}
\bibliography{references}

\end{document}